\documentclass{aastex63}

\received{}
\revised{}
\accepted{\today}
\submitjournal{AAS Journals}

\shorttitle{Pluto--Charon Sonata V}
\shortauthors{Kenyon \& Bromley}

\usepackage{epsf,amssymb,amsmath}
\usepackage{wrapfig}
\def\nbody{$n$-body}

\def\deg{\ifmmode {^\circ}\else {$^\circ$}\fi}
\def\degree{\ifmmode {^\circ}\else {$^\circ$}\fi}
\def\mum{\ifmmode {\rm \,\mu {\rm m}}\else $\rm \,\mu {\rm m}$\fi}

\def\inch{\ifmmode ^{\prime \prime}\else $^{\prime \prime}$\fi}
\def\gs{\ifmmode {{\rm g~s^{-1}}}\else ${\rm g~s^{-1}}$\fi}
\def\msunyr{\ifmmode {M_{\odot}~{\rm yr^{-1}}}\else $M_{\odot}~{\rm yr^{-1}}$\fi}
\def\msun{\ifmmode {M_{\odot}}\else $M_{\odot}$\fi}
\def\rsun{\ifmmode {R_{\odot}}\else $R_{\odot}$\fi}
\def\lsun{\ifmmode {L_{\odot}}\else $L_{\odot}$\fi}
\def\mstar{\ifmmode {M_{\star}}\else $M_{\star}$\fi}
\def\rstar{\ifmmode {R_{\star}}\else $R_{\star}$\fi}
\def\tstar{\ifmmode {T_{\star}}\else $T_{\star}$\fi}
\def\lstar{\ifmmode {L_{\star}}\else $L_{\star}$\fi}
\def\mwd{\ifmmode {M_{wd}}\else $M_{wd}$\fi}
\def\rwd{\ifmmode {R_{wd}}\else $R_{wd}$\fi}
\def\twd{\ifmmode {T_{wd}}\else $T_{wd}$\fi}
\def\lwd{\ifmmode {L_{wd}}\else $L_{wd}$\fi}
\def\md{\ifmmode {M_d}\else $M_d$\fi}
\def\ld{\ifmmode {L_d}\else $L_d$\fi}
\def\ad{\ifmmode A_d\else $A_d$\fi}
\def\ldlwd{\ifmmode L_d / L_{wd}\else $L_d / L_{wd}$\fi}
\def\ldlstar{\ifmmode L_d / L_\star\else $L_d / L_{\star}$\fi}
\def\lxlstar{\ifmmode L_X / L_\star\else $L_X / L_{\star}$\fi}
\def\rearth{\ifmmode {\rm R_{\oplus}}\else $\rm R_{\oplus}$\fi}
\def\mearth{\ifmmode {\rm M_{\oplus}}\else $\rm M_{\oplus}$\fi}
\def\qc{\ifmmode Q_c\else $Q_c$\fi}
\def\qdstar{\ifmmode Q_D^\star\else $Q_D^\star$\fi}
\def\rt{\ifmmode r_t\else $r_t$\fi}
\def\vc{\ifmmode v_c\else $v_c$\fi}
\def\vsqd{\ifmmode v^2 / Q_D^\star\else $v^2 / Q_D^\star$\fi}
\def\kms{\ifmmode {\rm km~s^{-1}}\else $\rm km~s^{-1}$\fi}
\def\masyr{\ifmmode {\rm mas~yr^{-1}}\else $\rm mas~yr^{-1}$\fi}
\def\ms{\ifmmode {\rm m~s^{-1}}\else $\rm m~s^{-1}$\fi}
\def\vrel{\ifmmode v_{rel}\else $v_{rel}$\fi}
\def\mdot{\ifmmode \dot{M}\else $\dot{M}$\fi}
\def\mdotz{\ifmmode \dot{M}_0\else $\dot{M}_0$\fi}
\def\mesc{\ifmmode m_{esc}\else $m_{esc}$\fi}
\def\rmin{\ifmmode r_{min}\else $r_{min}$\fi}
\def\rmax{\ifmmode r_{max}\else $r_{max}$\fi}
\def\xmax{\ifmmode x_{max}\else $x_{max}$\fi}
\def\mmin{\ifmmode m_{min}\else $m_{min}$\fi}
\def\mmax{\ifmmode m_{max}\else $m_{max}$\fi}
\def\rmind{\ifmmode r_{min,d}\else $r_{min,d}$\fi}
\def\rmaxd{\ifmmode r_{max,d}\else $r_{max,d}$\fi}
\def\mmaxd{\ifmmode m_{max,d}\else $m_{max,d}$\fi}
\def\mura{\ifmmode \mu_{RA}\else $\mu_{RA}$\fi}
\def\mudec{\ifmmode \mu_{Dec}\else $\mu_{Dec}$\fi}
\def\vrad{\ifmmode v_{rad}\else $v_{rad}$\fi}
\def\qz{\ifmmode q_{0}\else $q_{0}$\fi}
\def\qi{\ifmmode q_{i}\else $q_{i}$\fi}
\def\ql{\ifmmode q_{l}\else $q_{l}$\fi}
\def\qs{\ifmmode q_{s}\else $q_{s}$\fi}
\def\vhill{\ifmmode v_H\else $r_H$\fi}
\def\rhill{\ifmmode r_H\else $r_H$\fi}
\def\Rhill{\ifmmode R_H\else $R_H$\fi}
\def\rbrk{\ifmmode r_{brk}\else $r_{brk}$\fi}
\def\rdamp{\ifmmode r_{damp}\else $r_{damp}$\fi}
\def\rin{\ifmmode r_{in}\else $r_{in}$\fi}
\def\rout{\ifmmode r_{out}\else $r_{out}$\fi}
\def\tin{\ifmmode t_{in}\else $t_{in}$\fi}
\def\tout{\ifmmode t_{out}\else $t_{out}$\fi}
\def\ain{\ifmmode a_{in}\else $a_{in}$\fi}
\def\aout{\ifmmode a_{out}\else $a_{out}$\fi}
\def\r0{\ifmmode r_{0}\else $r_{0}$\fi}
\def\R0{\ifmmode R_{0}\else $R_{0}$\fi}
\def\m0{\ifmmode m_{0}\else $m_{0}$\fi}
\def\mone{\ifmmode m_{1}\else $m_{1}$\fi}
\def\mtwo{\ifmmode m_{2}\else $m_{2}$\fi}
\def\atwo{\ifmmode a_{2}\else $a_{2}$\fi}
\def\etwo{\ifmmode e_{2}\else $e_{2}$\fi}
\def\mf{\ifmmode m_{f}\else $m_{f}$\fi}
\def\af{\ifmmode a_{f}\else $a_{f}$\fi}
\def\ef{\ifmmode e_{f}\else $e_{f}$\fi}
\def\M0{\ifmmode M_{0}\else $M_{0}$\fi}
\def\amax{\ifmmode a_{max}\else $a_{max}$\fi}
\def\a0{\ifmmode a_{0}\else $a_{0}$\fi}
\def\e0{\ifmmode e_{0}\else $e_{0}$\fi}
\def\v0{\ifmmode v_{0}\else $v_{0}$\fi}
\def\xm{\ifmmode x_{m}\else $x_{m}$\fi}
\def\ag{\ifmmode A_{G}\else $A_{G}$\fi}
\def\av{\ifmmode A_{V}\else $A_{V}$\fi}
\def\ebmv{\ifmmode E_{B-V}\else $E_{B-V}$\fi}
\def\sigz{\ifmmode \Sigma_0\else $\Sigma_0$\fi}
\def\ergg{\ifmmode {\rm erg~g^{-1}}\else ${\rm erg~g^{-1}}$\fi}
\def\ergs{\ifmmode {\rm erg~s^{-1}}\else ${\rm erg~s^{-1}}$\fi}
\def\gyr{\ifmmode {\rm g~yr^{-1}}\else ${\rm g~yr^{-1}}$\fi}
\def\cms{\ifmmode {\rm cm~s^{-1}}\else ${\rm cm~s^{-1}}$\fi}
\def\gcms{\ifmmode {\rm g~cm^{-2}}\else $\rm g~cm^{-2}$\fi}
\def\gcmc{\ifmmode {\rm g~cm^{-3}}\else $\rm g~cm^{-3}$\fi}
\def\atil{\ifmmode {\tilde{a}}\else $\tilde{a}$\fi}
\def\ttil{\ifmmode {\tilde{t}}\else $\tilde{t}$\fi}
\def\sqrttt{\ifmmode {\tilde{t}^{1/2}}\else $\tilde{t}^{1/2}$\fi}

\def\ty2{{\it Tycho-2}}

\def\2mass{{\it 2MASS}}

\def\orch{{\it Orchestra}}
\def\nh{{\it New Horizons}}

\def\pc{Pluto--Charon}

\def\fs{\ifmmode f_S\else $f_S$\fi}
\def\ms{\ifmmode m_S\else $m_S$\fi}
\def\mp{\ifmmode m_P\else $m_P$\fi}
\def\mc{\ifmmode m_C\else $m_C$\fi}
\def\mh{\ifmmode m_H\else $m_H$\fi}
\def\mk{\ifmmode m_K\else $m_K$\fi}
\def\mn{\ifmmode m_N\else $m_N$\fi}
\def\rp{\ifmmode r_P\else $r_P$\fi}
\def\rc{\ifmmode r_C\else $r_C$\fi}
\def\apc{\ifmmode a_{PC}\else $a_{PC}$\fi}
\def\mpc{\ifmmode m_{PC}\else $m_{PC}$\fi}
\def\epc{\ifmmode e_{PC}\else $e_{PC}$\fi}
\def\rgc{\ifmmode r_{GC}\else $r_{GC}$\fi}
\def\qgc{\ifmmode q_{GC}\else $q_{GC}$\fi}
\def\Qgc{\ifmmode Q_{GC}\else $Q_{GC}$\fi}
\def\ag{\ifmmode a_{g}\else $a_{g}$\fi}
\def\eg{\ifmmode e_{g}\else $e_{g}$\fi}
\def\ageo{\ifmmode a_{geo}\else $a_{geo}$\fi}
\def\egeo{\ifmmode e_{geo}\else $e_{geo}$\fi}
\def\efree{\ifmmode e_{free}\else $e_{free}$\fi}
\def\ebin{\ifmmode e_{bin}\else $e_{bin}$\fi}
\def\abin{\ifmmode a_{bin}\else $a_{bin}$\fi}

\begin{document}

\title{A Pluto--Charon Sonata V. Long-term Stability of the HST State Vector}

\correspondingauthor{Scott J. Kenyon}
\email{skenyon@cfa.harvard.edu}

\author[0000-0003-0214-609X]{Scott J. Kenyon}
\affil{Smithsonian Astrophysical Observatory,
60 Garden Street,
Cambridge, MA 02138, USA}

\author[0000-0001-7558-343X]{Benjamin C. Bromley}
\affil{Department of Physics \& Astronomy,
University of Utah, 201 JFB,
Salt Lake City, DC 20006, USA}

\begin{abstract}
We analyze a new set of $\sim$ 275 \nbody\ calculations designed to place limits on the masses of the small circumbinary satellites in the \pc\ system. Together with calculations reported in previous papers, we repeat that a robust upper limit on the total mass of the four satellites is $\approx 9.5 \times 10^{19}$~g. For satellite volumes derived from \nh, this mass limit implies a robust upper limit on the bulk densities of Nix and Hydra, $\lesssim 1.7$~\gcmc, that are comparable to the bulk density of Charon. Additional calculations demonstrate that satellite systems with mass $\lesssim 8.25 \times 10^{19}$~g are robustly stable over the current age of the Sun. The bulk densities of Nix and Hydra in these lower mass systems are clearly smaller than the bulk density of Charon. These new \nbody\ results enable accurate measurements of eccentricity and inclination for Nix, Kerberos, and Hydra that agree well with orbital elements derived from numerical calculations with new HST and \nh\ state vectors. With these new state vectors, Styx has a 37\% larger eccentricity and an 85\% smaller inclination, which makes it more prone to gravitational perturbations from Nix.
\end{abstract}



\section{Introduction} \label{sec: intro}

Over the last two decades, space observations have added new insights into the physical properties of the dwarf planet Pluto \citep[e.g.,][]{stern2018,stern2021}. From 2005--2012, {\it Hubble Space Telescope} (HST) images revealed four small circumbinary satellites \citep{weaver2006,showalter2011,showalter2012}. Astrometric analyses of these and other HST data demonstrate that the orbits of the central \pc\ binary and the satellites are nearly circular and lie within a common plane \citep{buie2006,tholen2008,brozovic2015,showalter2015,giuppone2022}.  Spectacular observations acquired during the \nh\ flyby 
confirm that the satellites orbit within a common plane \citep{porter2023,brozovic2024} and tumble with approximate rotation periods of 0.43~d to 5.31~d \citep{showalter2015,weaver2016}.
All of the satellites are irregularly shaped and highly reflective, with characteristic radii $\sim$ 5~km for Styx and Kerberos and $\sim$ 20~km for Nix and Hydra \citep{weaver2016,cook2018}. More detailed analyses of the shapes suggest equivalent spherical radii of $\approx$ 18~km for Nix/Hydra, $\approx$ 6~km for Kerberos, and $\approx$ 5~km for Styx \citep{porter2021}. Although smaller satellites could exist slightly inside the orbit of Styx and outside the orbit of Hydra \citep{kb2019a}, there are no $\gtrsim$ 2~km satellites and a negligible amount of dust between the orbits of Styx and Hydra \citep{weaver2016,lauer2018}.

In several previous studies, we have used detailed \nbody\ calculations to place limits on the masses of individual satellites and on the combined mass of the four small satellites \citep[see also][]{youdin2012}. \citet{kb2019b} demonstrate that a `heavy' satellite system with individual satellite masses similar to those estimated in \citet{brozovic2015} and a total system mass $M_{SNKH} \sim 1.15 \times 10^{20}$~g is unstable on time scales $\lesssim$ 1~Gyr. Setting the bulk densities of Styx and Kerberos to $\rho_{SK} \sim$ 1~\gcmc\ reduces the system mass to a `nominal' mass $M_{SNKH} \sim 9.5 \times 10^{19}$~g. 
For an ensemble of calculations with masses of $f =$ 1--100 times the nominal mass, systems with $f \gtrsim$ 1.25 are generally unstable; systems with $f \approx$ 1 are often stable on 4~Gyr time scales \citep{kb2022}.
This result implies fairly low mass densities for Nix and Hydra, $\rho_{NH} \lesssim$ 1.4--1.5~\gcmc.

Here we conclude our study of the HST state vector for the \pc\ satellite system with a long-term stability analysis based on a completely new suite of \nbody\ calculations that consider satellite masses of zero and 0.5--0.875 times the nominally stable mass derived in \citet{kb2022}. This mass range is consistent with results from analyses of HST + \nh\ astrometric data \citep{porter2023,brozovic2024}. The new \nbody\ calculations demonstrate that all systems with a total mass $M_{SNKH} \lesssim 8.3 \times 10^{19}$~g ($f \lesssim$ 0.875) are completely stable over the 4.5~Gyr age of the solar system and provide new estimates of orbital eccentricity \efree\ and inclination $\imath$ for a stable system of four small circumbinary satellites in the \pc\ system. All satellites have a free eccentricity, \efree\ $\lesssim 6 \times 10^{-3}$, and orbital inclination, $\imath \lesssim 8 \times 10^{-3}$. Aside from Styx, these estimates for \efree\ and $\imath$ agree well with previous analyses of the Keplerian (osculating) orbital elements \citep{brozovic2015,showalter2015}. For Styx, the results discussed here account for orbital oscillations due to the central binary and place Styx on a more circular orbit than previous studies.  

Using a new set of \nbody\ calculations with the HST + \nh\ state vectors \citep{porter2023,brozovic2024}, we infer the median semimajor axis $a$, \efree, and $\imath$ for each satellite over a 100~Myr interval. Results for Nix, Kerberos, and Hydra are very close to estimates using the HST state vector. Although Styx has the same semimajor axis for all three state vectors, its \efree\ ($\imath$) is 37\% larger (85\% smaller) with the new state vectors. Although definite conclusions regarding the stability of the HST + \nh\ state vectors await the conclusion of a suite of \nbody\ calculations over 3--4.5~Gyr, we discuss possible implications for the stability of a Styx with a larger \efree\ and a smaller $\imath$ than in calculations with the HST state vector.

In the following sections, we describe the numerical procedures and the initial conditions for each calculation (\S\ref{sec: calcs}), the long-term stability and orbital properties of satellites in the \nbody\ calculations (\S\ref{sec: results}), and the significance of these results (\S\ref{sec: disc}). We conclude with a brief summary  (\S\ref{sec: summary}).

\section{Calculations} \label{sec: calcs}

\subsection{Procedures}

As in \citet{kb2019a,kb2019b,kb2019c,kb2022}, we perform numerical calculations with the gravitational \nbody\ code \orch, which uses a sixth-order accurate symplectic method in the center-of-mass frame to integrate the orbits of Pluto, Charon, and the four smaller satellites in response to their mutual gravitational interactions \citep{yoshida1990,wisdom1991,saha1992,bk2006,bk2020,kb2021a}. The calculations do not include the gravity of the Sun \citep[e.g.,][]{brozovic2015,michaely2017}, radiation pressure \citep[e.g.,][]{poppe2011,spahn2019}, or tidal dissipation \citep[e.g.,][]{quill2017}, which have negligible impact on the long-term dynamical evolution of satellite orbits. \citet{kb2022} describes additional details of the numerical procedures.

Throughout the \nbody\ calculations, \orch\ records the 6D cartesian phase space variables and other dynamical quantities at the end of selected time steps.  Over total integration times as long as 4.5~Gyr, a typical calculation has 30,000 to more than 100,000 `snapshots' at machine precision. On the NASA `discover' cluster, 24~hr integrations on a single processor advance the satellite system $\sim$ 5~Myr.  We perform 28--48 calculations per node, with each satellite system evolving on one of the 28--48 cores per node.  To derive results for as many sets of initial conditions as possible, the suite of simulations uses several nodes each day. In this way, each system advances $\sim$ 150~Myr per month.

\begin{deluxetable}{lccccccc}
\tablecolumns{8}
\tabletypesize{\footnotesize}
\tablenum{1}
\tablecaption{Adopted Masses and Initial Conditions\tablenotemark{\footnotesize \rm a}}
\tablehead{
\colhead{Satellite} &
\colhead{Mass (g)} &
  \colhead{$x$ (km)} &
  \colhead{$y$ (km)} &
  \colhead{$z$ (km)} &
  \colhead{$v_x$ (km~s$^{-1}$)} &
  \colhead{$v_y$ (km~s$^{-1}$)} &
  \colhead{$v_z$ (km~s$^{-1}$)}
}
\label{tab: init}
\startdata
Pluto & $1.303 \times 10^{25}$ & -157.8121679944 & -456.7988459683 & -2071.4067337364 & -0.0177032091 & -0.0158015359 & 0.0048362971 \\
Charon & $1.586 \times 10^{24}$ & 1297.1743847853 &  3752.6022617472 & 17011.9058384535 & 0.1453959509 & 0.1297771902 & -0.0397230040 \\
Styx & $6 \times 10^{17}$ & -30572.8427772584 & -26535.8134344897 & 12311.2908958766 & 0.0232883189 & 0.0427977975 & 0.1464990284 \\
Nix & $4.5 \times 10^{19}$ & 9024.3487802378 & 15210.7370165008 & 45591.7573572213 & 0.1004334400  &  0.0865524814 & -0.0479498746 \\
Kerberos & $9 \times 10^{17}$ & 23564.2070250521 & 28380.0399507624 & 44578.0258218278 & 0.0792537026 & 0.0630220100 & -0.0817084451 \\
Hydra & $4.8 \times 10^{19}$ & -43331.3261132443 & -43628.4575945387 & -20506.5419357332 & -0.0374001038 & -0.0184905611 & 0.1157937283 \\
\enddata
\tablenotetext{\rm a}{The 6D phase spaxce coordinates correspond to the Pluto-2 state vector in \citet{kb2019b}. The nominal masses for Nix and Hydra are 3--6 $\sigma$ (3--4 $\sigma$) larger than the best estimates of \citet{porter2023} \citep{brozovic2024}.}
\end{deluxetable}

\subsection{Initial Conditions}

All calculations begin with the 6D cartesian space state vector `Pluto-2' derived from HST data \citep{brozovic2015} as discussed in \citet{kb2019b} (Table~\ref{tab: init}). Adopting nominal masses for Styx, \ms\ = 0.6; Nix, \mn\ = 45; Kerberos, \mk\ = 0.9; and Hydra, \mh\ = 48 in units of $10^{18}$~g (Table~\ref{tab: init}), we multiply the nominal masses for each satellite by a factor $f = n (1 + \delta)$, where $n$ is an integer or simple fraction (e.g., 0.5, 0.625, 0.75, 0.875, 1.00, 1.25 or 1.5) and $\delta$ is a small real number in the range $-$0.01 to 0.01. For a suite of calculations with similar $f$, $n$ and $\delta$ are  {\it the same for all satellites}. Sometimes, we multiply the mass of a single satellite by a factor $f_i$ and set the masses of the remaining satellites at their nominal masses. To avoid confusion, we use $f$ as a marker for calculations where we multiply the masses of all satellites by a common factor and $f_i$ (where $i$ = `S' for Styx,`N' for Nix, `K' for Kerberos, and `H' for Hydra) as markers where 1--2 satellites have masses that differ from the nominal masses. In some calculations, we set $f_S = f_K$ = 1.5, 2, or 3 and then multiply masses for all four satellites by a common $f$. In these models, Styx and Kerberos have masses $f_S \times f$ larger than their nominal masses. For systems where all satellites have the same $f$, the $\delta$ term allows measurement of a range of lifetimes for systems with identical initial positions and velocities and nearly identical masses. 

Within the ensemble of \nbody\ calculations, physical collisions are exceedingly rare \citep{kb2022}. Adopted sizes for the satellites have no impact on collision outcomes. System stability depends only on adopted satellite masses. 

We define the system lifetime $\tau_i$ as the evolution time between the start of a calculation and the moment when one of the satellites is ejected beyond the \pc\ Hill sphere with an outward velocity that exceeds the local escape velocity, e.g. $v^2 > 2 ~ G ~ (\mp + \mc) / R_{H, PC}$. The uncertainty of the ejection time is negligible.  Among the $M$ calculations with nearly identical starting conditions, we adopt $\tau_m$ -- the median of $M$ different $\tau_i$ -- as the lifetime of the system. For fixed $f$, the range in $\tau_i$ is $\sim$ a factor of 3--100. 

As stated in Table~\ref{tab: init}, the nominal masses used here and in previous papers lie outside the 3-$\sigma$ mass limits derived in recent analyses of HST + \nh\ data \citep{porter2023,brozovic2024}. New numerical calculations with $f$ = 0.5--0.625 ($f$ = 0.75--0.875) have masses similar to (within 2--3-$\sigma$ of) current best estimates. Thus, the long-term evolution of new calculations with $f \lesssim$ 0.875 provide insight into the long-term stability of the new state vectors.

\subsection{Analysis}
\label{sec: analysis}

To analyze the \nbody\ calculations, we follow \citet{kb2022} and define four sets of orbital elements. Deriving the Keplerian elements $a_K$ and $e_K$ from the energy equation
\begin{equation}
\label{eq: akep}
v^2 = GM \left ( \frac{2}{r} ~ - ~ \frac{1}{a_K} \right ) 
\end{equation}
and the specific relative angular momentum, $h$,
\begin{equation}
\label{eq: ekep}
h^2 = GM ~ (1 ~-~ e_K^2) ~ a_K ~
\end{equation}
yields the pericenter $q = a (1 - e)$ and apocenter $Q = a (1 + e)$ of an orbit. 

The set of geometric elements rely on the distance of closest ($R_{min}$) and farthest ($R_{max}$) distances from the barycenter. With $R_{min}$ and $R_{max}$ as analogs of $q$ and $Q$, basic geometric relations for $a$ and $e$ follow
\citep[e.g.,][]{sutherland2019,kb2022}:
\begin{equation}
\label{eq: ag}
a_g = (R_{max} + R_{min}) / 2 
\end{equation}
and
\begin{equation}
\label{eq: eg}
e_g = (R_{max} - R_{min}) / (R_{max} ~ + ~ R_{min}) ~ .
\end{equation}
These measurements require some care to sample a single circumbinary orbit well 
or to collect a sufficient number of random snapshots over many circumbinary orbits.  

\citet{bk2020} derive an improved set of geometric elements based on the \citet{lee2006} theory derived from solutions to the restricted three-body problem. Identifying corrections to $R_{min}$ and $R_{max}$ appropriate for satellites orbiting the center of mass of a circular binary system, \citet{bk2020} estimate the semimajor axis \ageo\ and eccentricity \egeo. \citet{kb2022} describe the approach in more detail and apply it to numerical calculations of the \pc\ satellites.

In addition to \ageo\ and \egeo, \citet{bk2020} develop an iterative solution for the semimajor axis and eccentricity in the context of \citet{lee2006} theory from one 6D state vector for each satellite. Formally, the estimate for the semimajor axis is analogous to the radius of the guiding center \rgc\ in the \citet{lee2006}; similarly, the eccentricity is equivalent to the free eccentricity \efree. \citet{bk2020} show that the iterative solution for \rgc\ and \efree\ agrees with the analysis of \citet{woo2020}, who infer orbital elements in the context of \citet{lee2006} theory from an FFT of satellite positions throughout an orbit. 

\subsection{System Stability}
\label{sec: stability}

In \S\ref{sec: results}, we explore the dynamics of circumbinary satellite systems as a function of the mass factors $f$ and $f_i$. The numerical calculations show that systems with $f \le$ 0.875 are uniformly stable; those with $f \ge$ 1.5 are always unstable. For $f \approx$ 1--1.25, some systems are stable; others are not. 

To understand this behavior, it is helpful to define the mutual Hill radius $R_{H, ij} = ((m_i + m_j) / 3 (m_P + m_C))^{1/3} a_i$, where $m_i$ and $m_j$ ($a_i$ and $a_j$) are the masses (semimajor axes) of a pair of satellites. We express the differences in the semimajor axes (e.g., $a_S - a_N$) in terms of $R_H$, $a_i - a_j = K R_{H, ij}$. Thus, $K_{SN}$ = 12 for Styx--Nix, $K_{NK}$ = 16 for Nix--Kerberos, and $K_{KH}$ = 10 for Kerberos--Hydra.  For circular and coplanar orbits as in the \pc\ system, numerical calculations require $K \gtrsim$ 8--10 for stablility \citep[e.g.,][]{wisdom1980,petit1986,gladman1993,chambers1996,deck2013,fang2013,fabrycky2014,kratter2014,mahajan2014,pu2015,morrison2016,obertas2017,weiss2018,sutherland2019,volk2024}. These results suggest that the four small satellites with their nominal masses are approximately stable. Decreasing (increasing) the mass should lead to more (less) stable systems. With the Kerberos--Hydra pair having the smallest $K$, Kerberos is the most likely satellite to be ejected from an unstable system. The calculations described below are designed to test this analysis. 

Throughout a long numerical sequence, satellites must also avoid resonances, semimajor axes where the orbital period is an integer multiple of the \pc\ orbital period. Satellites in resonance are unstable and have short lifetimes in the satellite system \citep[e.g.,][]{ward2006,youdin2012,bk2015b,walsh2015}. Starting from the \citet{brozovic2015} state vector, all four satellites have nearly circular orbits, $e \lesssim$ 0.006, that do not cross nearby resonances. Orbits with massless satellites are therefore stable over the expected lifetime of the solar system. When satellites have mass, gravitational interactions excite satellite orbits, which increases $e$ and perhaps $\imath$. Kerberos and Styx are most at risk of crossing orbital resonances with the central binary. Kerberos avoids the 5:1 resonance as long as $e \lesssim$ 0.01 \citep{kb2022}. When Styx maintains an orbit with $e \lesssim$ 0.033, it dodges the 3:1 resonance. 

In \S\ref{sec: lifetimes}, we consider the lifetimes of these systems as a function of $f$ and $f_S$. When $f \lesssim$ 0.875 ($\gtrsim$ 1.5), \pc\ satellite systems are uniformly stable (unstable). For intermediate $f$, some systems maintain stability for 4.5~Gyr; others eject at least one satellite. We then explore explanations for the curious dichotomy in satellite ejections, where ejections of Styx (Kerberos) are much more likely when $f =$ 1 ($f \ge$ 1.25; \S\ref{sec: ejections}). In \S\ref{sec: orbs}, we compare estimates of $e$ and $\imath$ for stable systems derived from the \nbody\ calculations. Among systems with $f \lesssim$ 0.875, the estimates for $e$ and $\imath$ have very little scatter. In \S\ref{sec: newstate}, we compare the orbital elements derived for the HST state vector with those from preliminary calculations with the new state vectors derived from HST + \nh\ data \citep{porter2023,brozovic2024}.

\section{Results} \label{sec: results}

\subsection{System Lifetimes}
\label{sec: lifetimes}

Table~\ref{tab: lifetimes} summarizes several aspects of the full set of calculations with $f$ = 0.5--1.5 and $f_S = f_K$ = 1--3. For $f \gtrsim$ 1.50 (1.25), (almost) all systems are unstable on time scales ranging from a minimum of $\sim$ 140~Myr to a maximum of $\sim$ 4.4~Gyr. These evolutionary sequences follow the same trend. Initially, Styx and Kerberos have nearly constant $a$ and $e$. However, the motion of Styx or Kerberos slowly begins to deviate from the original, nearly-circular orbit. One satellite then crosses the 3:1 (Styx) or the 5:1 (Kerberos) resonance with the central binary. After resonance crossing, the orbit of Styx (Kerberos) is excited to larger $e$ and $\imath$; it then crosses the orbit of Nix (Hydra) and then the innermost stable orbit of the \pc\ binary. Once in the vicinity of \pc, the satellite is rapidly ejected from the system.

\begin{deluxetable}{cccccccc}
\tablecolumns{8}
\tablewidth{15cm}
\tabletypesize{\normalsize}
\tablenum{2}
\tablecaption{Lifetimes for Model Satellite Systems\tablenotemark{\footnotesize \rm a}}
\tablehead{
\colhead{~~~$f$~~~} &
\colhead{~~~~~$f_S, f_K$~~~~~} &
\colhead{~~~~$N_0$~~~~} &
\colhead{~~~~$N_e(S)$~~~~} &
\colhead{~~~~$N_e(K)$~~~~} &
\colhead{~~~~~$\tau_{min}$~~~~~} &
\colhead{~~~~~$\tau_m$~~~~~} &
\colhead{~~~~~$\tau_{max}$~~~~~}
}
\label{tab: lifetimes}
\startdata
0.000 & 1.0 & 7 & 0 & 0 & 9.65 & 9.65 & 9.65 \\
0.500 & 1.0 & 14 & 0 & 0 & 9.65 & 9.65 & 9.66 \\
0.625 & 1.0 & 14 & 0 & 0 & 9.65 & 9.65 & 9.65 \\
0.750 & 1.0 & 14 & 0 & 0 & 9.65 & 9.65 & 9.65 \\
0.875 & 1.0 & 14 & 0 & 0 & 9.65 & 9.65 & 9.65 \\
1.000 & 1.0 & 16 & 8 & 1 & 8.76 & 9.61 & 9.65 \\
1.000 & 1.5 & 15 & 4 & 1 & 9.15 & 9.65 & 9.65 \\
1.000 & 2.0 & 19 & 6 & 2 & 9.23 & 9.65 & 9.65 \\
1.000 & 3.0 & 15 & 3 & 5 & 9.07 & 9.52 & 9.65 \\
1.250 & 1.0 & 32 & 8 & 22 & 8.71 & 9.24 & 9.65 \\
1.250 & 1.5 & 32 & 5 & 25 & 8.52 & 9.07 & 9.65 \\
1.250 & 2.0 & 29 & 9 & 19 & 8.58 & 9.11 & 9.65 \\
1.250 & 3.0 & 35 & 5 & 28 & 8.34 & 9.11 & 9.65\\
1.500 & 1.0 & 34 & 11 & 23 & 8.26 & 8.89 & 9.49\\
1.500 & 1.5 & 29 & 11 & 18 & 8.26 & 8.74 & 9.33\\
1.500 & 2.0 & 30 & 10 & 20 & 8.27 & 8.75 & 9.35\\
1.500 & 3.0 & 29 & 9 & 20 & 8.14 & 8.64 & 9.37 \\
\enddata
\tablenotetext{\rm a}{The columns list $f$, $f_{S, K}$, the number of completed calculations $N$, the number of calculations resulting in the ejection of Styx $N_e(S)$ or Kerberos $N_e(K)$, and the log of the minimum ($\tau_{min}$), median ($\tau_m$), and maximum ($\tau_{max}$) lifetimes (in yr) among the $N$ calculations for each combination of $f$ and $f_{S, K}$.
}
\end{deluxetable} 

Systems with $f$ = 1 exhibit two types of outcomes. In nearly half (30 out of 64) of the calculations, Styx or Kerberos is ejected on time scales ranging from  $\sim$ 600~Myr to $\sim$ 4.4~Gyr. The other half (34 out of 64) reach the end of the calculation at 4.5~Gyr. With a median lifetime comparable to the age of the solar system, the \citet{brozovic2015} state vector is marginally stable when the small \pc\ satellites have their nominal masses.

Lower mass satellite systems with $f$ = 0.500--0.875 and $f_S$ = 1 are uniformly stable over the current age of the Solar System. Including the 7 calculations with massless satellites, all 63 calculations with $f \lesssim$ 0.875 reach the 4.5~Gyr endpoint without a single ejection of Styx or Kerberos. Given the low masses of these systems, we are confident that calculations with $f \lesssim$ 0.875 and $f_S$ = 1.5--3.0 are also uniformly stable on 4.5~Gyr time scales.

\begin{figure}[t]
\begin{center}
\includegraphics[width=0.55\linewidth]{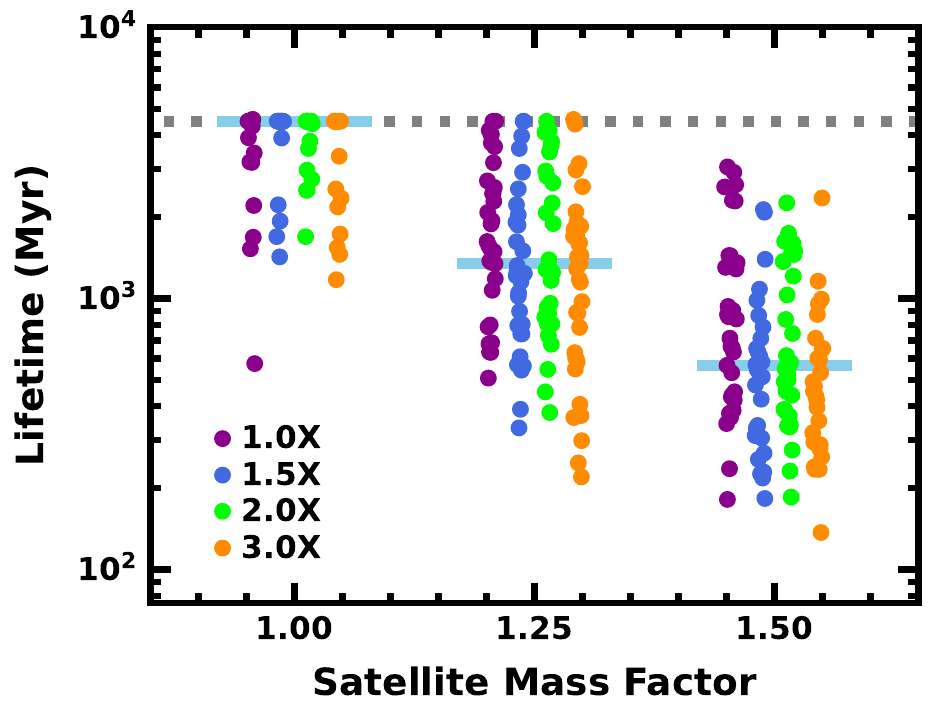}
\vskip -1ex
\caption{
\label{fig: all-life}
Lifetime of each model \pc\ satellite system as a function of the mass factor $f$ for calculations with (from left to right within each group) $f_{S, K}$ = 1.0 (purple points), 1.5 (blue points), 2.0 (green points), and 3.0 (orange points). The horizontal grey dotted line indicates the current age of the Sun. The light blue horizontal lines indicate the median lifetime of the full set of calculations with $f$ = 1.0, 1.25, or 1.5 and any $f_{S, K}$.
}
\end{center}
\end{figure}

Fig.~\ref{fig: all-life} shows the ensemble of system lifetimes $\tau$ for $f$ = 1.00, 1.25, and 1.50 and $f_S$ = 1.0--3.0. There is a clear trend of increasing $\tau$ with decreasing $f$. The single outlier in the distribution of lifetimes for $f = f_S$ = 1 shows the occasional impact of chaotic behavior where the early timing of resonance crossing leads to a rapid ejection compared to other calculations. 

In previous papers \citep{kb2019b,kb2022}, we considered whether the slightly larger masses of Styx and Kerberos in calculations with $f \ge$ 1.5 and $f_S$ = 3 shortened lifetimes relative to those with $f \ge$ 1.5 and $f_S$ = 1. With the data set for $f$ = 1.00--1.25 now complete, we revisit this issue to try to place some limits on the masses of Styx and Kerberos. For calculations with $f$ = 1.25 and 1.50, the median lifetimes as a function of $f_S$ show a clear trend: systems with larger $f_S$ have a shorter median lifetime than those with smaller $f_S$. The trend is less clear for calculations with $f$ = 1.00: the median lifetime for $f_S$ = 3 is shorter than those for smaller $f$, but median lifetimes for $f_S$ = 1.0, 1.5, and 2.0 are similar. There is also no trend in the number of calculations that led to the ejection of a satellite $N_e$ with \fs. Systems with \fs\ = 1.5 have the smallest fraction of ejections, while those the \fs\ = 1.0, 2.0, and 3.0 have similar ejection fractions. 

For a more quantitative approach, we examine results for the Python version of the Mann--Whitney--Wilcoxon rank-sum test, which ranks the lifetimes and uses a non-parametric technique to measure the probability that the two distributions are drawn from the same parent population. For $f$ = 1.5, adding new lifetimes from several calculations not included in the \citet{kb2022} data set does not change the results. The rank-sum tests return small probabilities, $p$ = 0.13 to less than 0.001, that the calculations with $f_S$ = 1 and $f_S$ = 1.5, 2.0, and 3.0 are drawn from the same parent population. The data for \fs\ = 1.0 and \fs\ = 1.5 (\fs\ = 3.0) are most likely (least likely) to be drawn from the same population. Results for $f$ = 1.25 follow this trend. The distributions of lifetimes for \fs\ = 1.0 and \fs\ = 1.5 are most likely ($p$ = 0.2) to be drawn from the same parent; lifetimes for \fs\ = 2.0 and \fs\ = 3.0 are least likely ($p$ = 0.007). 

The trends measured with the Mann--Whitney--Wilcoxon probabilities for $f$ = 1.25 and 1.50 do not continue for calculations with $f$ = 1.00. For any combination of \fs, $p \approx$ 0.25 to 0.66, which suggests that all of the lifetimes could be drawn from the same parent population. With smaller samples for each \fs, these results might be compromised by smaller number statistics. 

Using the Python version of the Komolgorov--Smirnov test yields nearly identical results as the Mann--Whitney--Wilcoxon test \citep[see also][]{kb2022}. For calculations with $f$ = 1.25 and 1.50, there is a clear trend that the probability that the samples are drawn from different parents is $p \le$ 0.001 for \fs\ = 1.0 and \fs\ = 3 and $p \approx$ 0.2--0.3 for \fs\ = 1.0 and either \fs\ = 1.5 or \fs\ = 2.0. For calculations with $f$ = 1.0, however, the tests return high probabilities, $p \gtrsim$ 0.2, that the samples are drawn from the same parent for any \fs.

We conclude that calculations with $f$ = 1.25 and 1.50 are so unstable on 4~Gyr time scales that small increases in the masses of Styx and Kerberos (e.g., \fs\ = 3.0) are sufficient to decrease system lifetimes significantly. For the marginally stable satellite systems with $f$ = 1, the statistical tests rely on small sample sizes, but they still suggest that increasing the masses of Styx and Kerberos by factors of 2--3 relative to their nominal masses has little or no impact on system lifetimes.

\subsection{Styx? Or Kerberos?}
\label{sec: ejections}

The numerical \nbody\ calculations described here and in \citet{kb2019b,kb2022} have a curious feature. When the mass factor is large ($f \ge 1.25$), ejections of Kerberos are $\gtrsim$ 2 times more frequent than ejections of Styx \citep[Table~\ref{tab: lifetimes}; see also Table 2 of][]{kb2022}. When $f \approx$ 1, however, unstable systems eject Styx 2--3 times more often than Kerberos. Understanding this dichotomy provides insights into circumbinary dynamics.

For the nominal masses and orbits, Kerberos appears easier to eject than Styx. For any pair of small satellites, Kerberos--Hydra has the smallest orbital separation in Hill units, $K_{KH}$ = 10. Although Styx is closer to the innermost stable orbit of the central binary, Styx--Nix has $K_{SN}$ = 12. As noted above, Kerberos (Styx) must have $e \lesssim$ 0.01 (0.033) to avoid crossing the 5:1 (3:1) orbital resonance. With $e_K \approx$ 0.004 and $e_S \approx$ 0.001, Kerberos requires much smaller increases in $e$ to cross a resonance. These factors could make Styx `safer' for all satellite masses. 

Small angle scattering provides some insight into the preference for Styx ejections when the satellites have their nominal masses and $f$ = 1 \citep[e.g.,][]{lin1979a,lin1979b,bk2011b}. With evolution times $\sim$ 1 Gyr (Figure~\ref{fig: all-life}), small deflections of Styx and Kerberos by their more massive companions dominate the growth of orbital eccentricity. Deflections are inversely proportional to the orbital separation in Hill units $K$ and occur on the synodic period, which is 107 days for Styx--Nix, 109 days for Nix--Kerberos, and 204 days for Kerberos--Hydra. With the smallest $K$, Kerberos receives the largest deflection every synodic period. However, Styx has only a somewhat larger $K$ (12 vs. 10); Nix deflects it twice as often as Hydra deflects Kerberos. The difference in synodic periods more than makes up for the somewhat weaker gravitational perturbations of Nix. Although Nix exerts a pull on Kerberos on time scales similar to its pull on Styx, the larger $K$ results in much weaker perturbations. When small angle deflections dominate the evolution, this analysis suggests Styx should be ejected more often than Kerberos as observed in the numerical calculations. 


As satellite masses (and $f$) grow, small angle scattering becomes less important. When $f$ = 2--3, $K_{KH} \approx$ 8--7 and $K_{SN}$ = 10--8.5. Based on a variety of numerical calculations, large perturbations of satellite orbits begin to dominate when $K \lesssim$ 7--8, leading to more rapid ejections \citep[e.g.,][]{wisdom1980,petit1986,gladman1993,chambers1996,deck2013,fang2013,fabrycky2014,kratter2014,mahajan2014,
pu2015,morrison2016,obertas2017,weiss2018,sutherland2019}. To evaluate the initial response of the satellites to increasing $f$, we consider the evolution of orbital elements on short time scales, $\sim$ 1000~yr, where we record the state vectors much more frequently than in the longer integrations described above. The finer sampling of these 1000~yr calculations provide more accurate eccentricity estimates over a single orbit than estimates derived from the less frequent sampling of the much longer integrations. For these estimates, we use the corrected geometric elements \ageo\ and \egeo\ and the estimates \rgc\ and \efree\ derived from the linear theory of \citet{bk2021}. 

\begin{figure}
    \centering
    \includegraphics[width=0.55\linewidth]{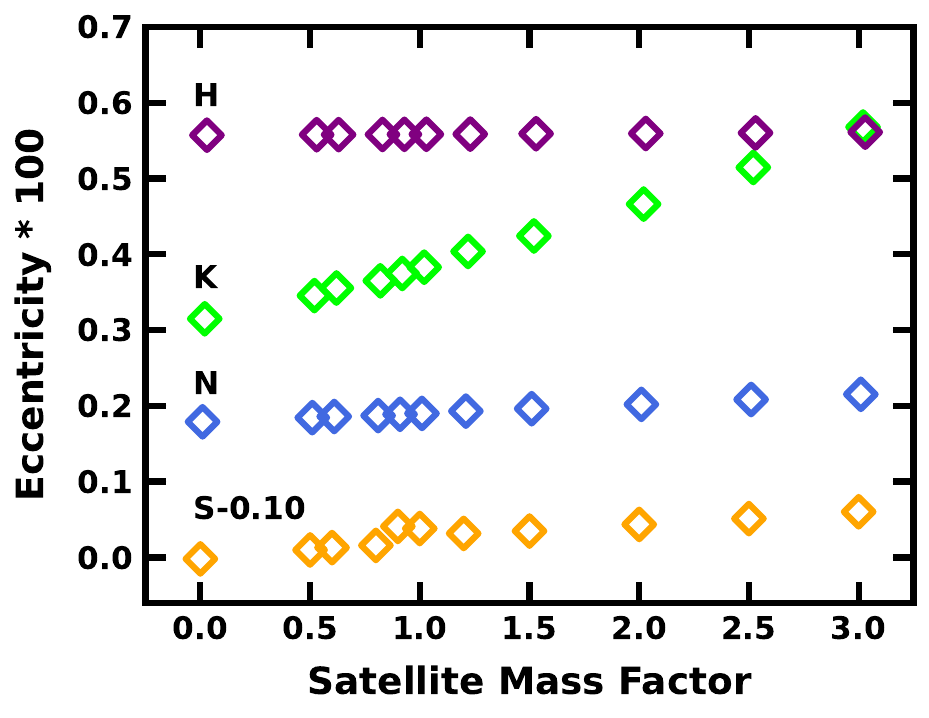}
    \caption{Eccentricity of the small satellites as a function of mass factor $f$ during a 1000 yr orbital integration. The symbols show the geometric eccentricity, corrected for ubiquitous influence of the central binary to highlight the free eccentricity. The most-circular estimates follow the trends shown in the figure. The values for Styx have been shifted as indicated for clarity.}
    \label{fig:evf}
\end{figure}

During 1000~yr, satellite orbits evolve on very short time scales, $\sim$ 1--10~yr, to a rough equilibrium state where the semimajor axis and eccentricity remain nearly constant until the long-term gravitational forces of Nix and Hydra begin to pull a satellite onto a path that leads to ejection. Figure~\ref{fig:evf} illustrates the typical satellite eccentricity as a function of the mass factor $f$ during this 1000~yr interval. Over the length of the calculations, eccentricity variations are much smaller than the sizes of the points.


Figure~\ref{fig:evf} shows a clear difference between the short-term evolution of the four satellites. For Nix and Hydra, the eccentricity for $f$ = 3 is only somewhat larger than the eccentricity for $f$ = 0--1. Clearly, Nix and Hydra react little to large changes in satellite masses. Styx and Kerberos behave differently. When $f$ = 0--1, the eccentricity for Styx grows by $\sim$ 33\%; for larger $f$, the growth in $e$ is much smaller. Thus, Styx reacts more to changes in the system mass when $f$ is small than when $f$ is large. In contrast, the eccentricity of Kerberos varies more strongly with $f$; the change in $e$ is smaller when $f$ = 0--1 than when $f$ = 1--3. 

The behavior of Styx and Kerberos in Figure~\ref{fig:evf} leads to a better understanding of the ejection frequency dichotomy. When $f \lesssim$ 1, neither satellite responds much to changes in the system mass. Small angle scattering then dominates and generally leads to ejections of Styx. When $f \gtrsim$ 1.25--1.5, Kerberos' orbit becomes much more excited than Styx' orbit on 1000~yr (and much longer) time scales. For Kerberos, the larger $e$ moves the orbit closer to the 5:1 resonance and reduces $K_{KH}$ ($K_{KN}$) near apocenter (pericenter). Kerberos then receives stronger perturbations from Hydra and Nix. With the same $e_S$ for any $f$, Styx receives relatively smaller perturbations compared to Kerberos as $f$ grows. Kerberos is then more
likely to be ejected than Styx, as the calculations show.

\subsection{Satellite Orbital Parameters}
\label{sec: orbs}

To derive orbital parameters for Styx, Nix, Kerberos, and Hydra, we compile $a_K$, $a_g$, \rgc, $a_{geo}$, $e_K$, $e_g$, \efree, $e_{geo}$, $\imath_K$, $\imath_g$, and $\imath_{free}$ for every snapshot in calculations with $f \le 1.25$ and lifetimes $\tau \gtrsim$ 4.5~Gyr. From the full set of snapshots in each calculation, we derive the median value and inter-quartile range of each orbital parameter. Within the set of median values and inter-quartile ranges for a particular $f$, the values are identical to within 0.001 in the semimajor axis or radius of the guiding center and within $10^{-5}$ for the eccentricities and inclinations.  

\begin{figure}[t]
\begin{center}
\includegraphics[width=0.55\linewidth]{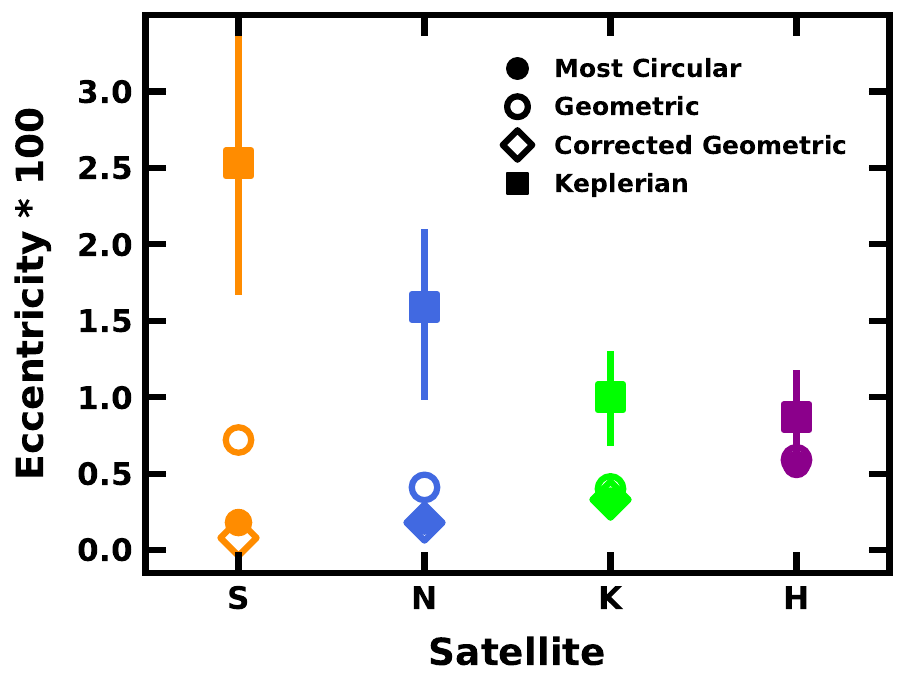}
\vskip -1ex
\caption{
\label{fig: allecc}
Median eccentricity for calculations with $f$ = 0.5 from the iterative most circular \citep[filled circles;][]{bk2021}, geometric (open circles), geometric with most circular corrections \citep[open diamonds;][]{bk2021}, and Keplerian (filled squares) approaches for each \pc\ satellite. Vertical lines indicate inter-quartile ranges for the Keplerian estimates. Inter-quartile ranges for the most circular and the two geometric approaches are smaller than the symbols. At large orbital separation, all four approaches yield similar estimates. For smaller orbital separations, the Keplerian (geometric) approach(es) diverges significantly (minimally) from the most circular estimate.
}
\end{center}
\end{figure}

Figure~\ref{fig: allecc} compares the four approaches (Keplerian, geometric, most circular correction to geometric, and most circular) for the full set of 14 calculations with $f$ = 0.5. Each point represents the median eccentricity for each satellite derived from the 4.5~Gyr of evolution for 14 calculations. Error bars indicate the inter-quartile range for the Keplerian estimates. Inter-quartile ranges for the other estimates are comparable to or smaller than the sizes of the points. The two geometric estimates have factor of 10--20 smaller inter-quartile ranges than the most circular estimates (Table~\ref{tab: elements}).

As discussed in \citet{bk2021} and \citet{kb2022}, the agreement between these estimates depends on the distance from the system barycenter. When this distance is large, a satellite on a nearly circular orbit feels a roughly constant gravitational force throughout its orbit. The four estimates of the eccentricity (and $a$ or \rgc) agree well. Hydra provides a good example. Here, the two geometric and the most circular approaches are nearly identical. The uncertainty as measured by the inter-quartile range is fairly small for all three estimates (Table~\ref{tab: elements}). The Keplerian estimate for $e$ is only somewhat larger than the other estimates. As the separation from the barycenter decreases, satellites feel a larger and larger variation in the gravitational force from the binary as a function of time. The four estimates then diverge. For Styx, the satellite experiences the largest radial motion due to the time-varying potential. The Keplerian estimates cannot compensate for this intrinsic motion of a nearly circular orbit around a binary and thus yield a much larger eccentricity than the other estimates. The simple geometric estimate compensates for the extra motion of Styx better, but it does not capture the motion as well as the corrected geometric estimate or the most circular estimate based on the analytical theory of \citet{lee2006}. Thus, we favor the corrected geometric estimates or the most circular estimates for \efree.

Table~\ref{tab: ecc} lists the average median value and inter-quartile range for each mass factor $f$. Among all calculations, the derived \efree\ is nearly independent of the mass factor $f$. Styx has the smallest \efree, 0.00165--0.00184; Hydra has the largest, 0.00549--0.00559. Kerberos has the largest range in \efree\ among all of the calculations, 0.00321--0.00363, and shows the clearest dependence on $f$: \efree\ is larger for larger mass factors. For all satellites, the variations in \efree\ throughout the calculations are small, ranging from $\lesssim 7 \times 10^{-4}$ for Styx and Kerberos to $\lesssim 2-3 \times 10^{-5}$ for Nix and Hydra. These variations are also remarkably stable with $f$. For Styx, Nix, and Hydra, the variations are similar for all $f$. The variations in \efree\ for Kerberos grow slowly with $f$. Clearly, the orbits are stable over 4.5 Gyr, which is $\sim 2.56 \times 10^{11}$ orbits of the central binary. 

\begin{deluxetable}{lcccc}
\tablecolumns{5}
\tablewidth{15cm}
\tabletypesize{\normalsize}
\tablenum{3}
\tablecaption{Satellite eccentricity as a function of satellite mass factor}
\tablehead{
\colhead{mass factor} &
\colhead{~~~~~~~~~Styx~~~~~~~~~} &
\colhead{~~~~~~~~~Nix~~~~~~~~~} &
\colhead{~~~~~~~~~Kerberos~~~~~~~~~} &
\colhead{~~~~~~~~~Hydra~~~~~~~~~}
}
\label{tab: ecc}
\startdata
0.000 & 0.183$\pm$0.076 & 0.193$\pm$0.020 & 0.321$\pm0$.008 & 0.559$\pm$0.003\\
0.500 & 0.178$\pm$0.073 & 0.194$\pm$0.020 & 0.323$\pm$0.013 & 0.555$\pm$0.003\\
0.625 & 0.178$\pm$0.073 & 0.194$\pm$0.020 & 0.324$\pm$0.016 & 0.554$\pm$0.004\\
0.750 & 0.177$\pm$0.072 & 0.195$\pm$0.020 & 0.324$\pm$0.019 & 0.553$\pm$0.004\\
0.875 & 0.174$\pm$0.069 & 0.195$\pm$0.019 & 0.336$\pm$0.023 & 0.552$\pm$0.005\\
1.000 & 0.188$\pm$0.072 & 0.196$\pm$0.019 & 0.347$\pm$0.028 & 0.551$\pm$0.006\\
1.250 & 0.174$\pm$0.057 & 0.196$\pm$0.019 & 0.325$\pm$0.067 & 0.549$\pm$0.007\\
\enddata
\tablecomments{Quoted values (errors) are 100 times the median \efree\ (inter-quartile range for the median \efree) derived from all calculations for each mass factor.} 
\end{deluxetable}

Table~\ref{tab: elements} summarizes the adopted elements for the four satellites based on calculations with $f$ = 0.5. To avoid confusion, Keplerian elements are not included. For all four satellites, \rgc, $a_{geo}$, and $a_g$ are very stable throughout a 4.5~Gyr integration. Variations in the orbital eccentricity and inclination are small for Nix, Kerberos, and Hydra. Styx has much larger variations. All four satellites have stable estimates of the inclination. 

Table~\ref{tab: elements} also illustrates some of the difficulties in deriving orbital elements. In the geometric approach, the semimajor axis and eccentricity depend on the path of each satellite about the central binary over many orbits. In some sense, the derived orbital path averages over fluctuations in the potential and is thus very stable through each calculation. In contrast, the most circular estimates are instantaneous measures of the orbit and thus have larger variations.

\begin{deluxetable}{lcccc}
\tablecolumns{5}
\tablewidth{15cm}
\tabletypesize{\normalsize}
\tablenum{4}
\tablecaption{Orbital Elements for Model Satellite Systems}
\tablehead{
\colhead{Parameter} &
\colhead{~~~~~~~~~Styx~~~~~~~~~} &
\colhead{~~~~~~~~~Nix~~~~~~~~~} &
\colhead{~~~~~~~~~Kerberos~~~~~~~~~} &
\colhead{~~~~~~~~~Hydra~~~~~~~~~}
}
\label{tab: elements}
\startdata
\rgc\ (\rp) & 35.640$\pm$0.004 & 40.961$\pm$0.002 & 48.590$\pm$0.003 & 54.460$\pm$0.005 \\
$a_g$ (\rp) & 35.761$\pm$0.002 & 41.002$\pm$0.001 & 48.604$\pm$0.002 & 54.468$\pm$0.001 \\
$a_{geo}$ (\rp) & 35.534$\pm$0.002 & 40.912$\pm$0.001 & 48.570$\pm$0.003 & 54.540$\pm$0.001 \\
100 * $e_{free}$ & 0.178$\pm$0.073 & 0.194$\pm$0.020 & 0.323$\pm$0.003 & 0.555$\pm$0.003 \\
100 * $e_g$ & 0.723$\pm$0.005 & 0.407$\pm$0.002 & 0.403$\pm$0.004 & 0.587$\pm$0.002 \\
100 * $e_{geo}$ & 0.084$\pm$0.006 & 0.184$\pm$0.003 & 0.330$\pm$0.005 & 0.551$\pm$0.002 \\
100 * $\imath_{free}$ & 0.446$\pm$0.025 & 0.039$\pm$0.001 & 0.778$\pm$0.003 & 0.536$\pm$0.001 \\
100 * $\imath_g$ & 0.477$\pm$0.001 & 0.039$\pm$0.000 & 0.780$\pm$0.001 & 0.536$\pm$0.000 \\
\enddata
\tablecomments{Quoted values (errors) are 100 times the median \efree\ (inter-quartile range for the median \efree) derived from all calculations for $f$ = 0.5. Inter-quartile ranges less than $5 \times 10^{-6}$ are quoted as 0.000.}
\end{deluxetable}

\begin{figure}[t]
\begin{center}
\includegraphics[width=0.55\linewidth]{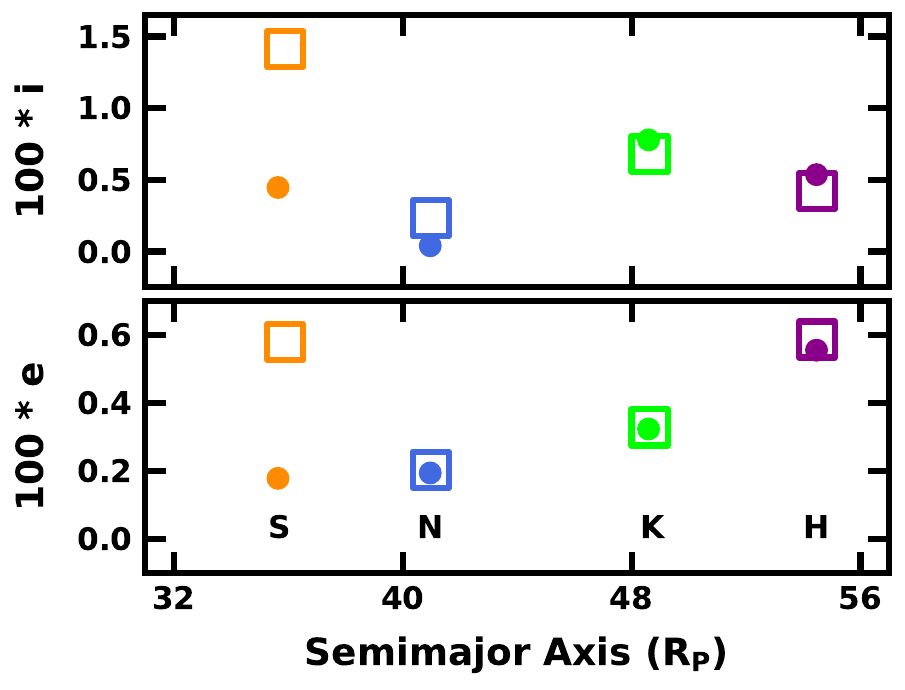}
\vskip -1ex
\caption{
\label{fig: ae}
Comparison of the median orbital elements inferred from the numerical calculations described in the text (filled circles) with elements derived from HST data \citep[large open squares;][]{showalter2015}. For Nix, Kerberos, and Hydra, elements from the numerical calculations agree very well with the HST-only elements. The numerical calculations derive much smaller estimates for the eccentricity and inclination of Styx. The lower panel includes an initial for each satellite at $e$ = 0. 
}
\end{center}
\end{figure}

To conclude this sub-section, Figure~\ref{fig: ae} compares the orbital elements derived from our \nbody\ calculations for $f$ = 0.5 (filled circles; Table~\ref{tab: elements}) with those derived from detailed Keplerian orbital fits to HST data that allow for apsidal precession and nodal regression \citep[open squares;][]{showalter2015}. Uncertainties from the \nbody\ calculations are similar to or smaller than the sizes of the symbols. For the HST measurements, the eccentricities for Nix, Kerberos, and Hydra are equally small; the fractional uncertainty for Styx is $\sim$ 20\% \citep{showalter2015}. In the lower panel, the two estimates for \rgc--$a$ and eccentricity match very well for Nix, Kerberos, and Hydra. The HST estimate for $e$ is somewhat larger than the \nbody\ estimate for Hydra. The \citet{showalter2015} eccentricity estimate for Styx is much larger than the \nbody\ estimate. Because \citet{showalter2015} derived $e$ from fitting an ellipse to a non-elliptical path around the binary, we expect this approach to result in a much larger $e$ for Styx.

The upper panel of Figure~\ref{fig: ae} compares estimates for semimajor axis and orbital inclination. Once again, the estimates agree well for Nix, Kerberos, and Hydra: the HST results are somewhat larger (smaller) for Nix (Kerberos and Hydra) than the \nbody\ estimates. Styx is an outlier; the \citet{showalter2015} estimate lies well above the \nbody\ estimate. We suspect this difference is another result of the difficulty in fitting ellipses to the orbit of Styx, when its path around the system barycenter is pushed and pulled by the time-varying potential of the binary.  

\subsection{Comparison with Recent State Vectors}
\label{sec: newstate}

While this study was underway, \citet{porter2023} and \citet{brozovic2024} announced state vectors and orbital elements from independent analyses of HST + \nh\ astrometric data for the \pc\ system. Both studies estimate $e$ and $\imath$ from osculating orbital elements measured from 1000~yr \nbody\ calculations of the full satellite system. These agree well with the osculating elements from the \nbody\ calculations based on the HST state vector reported here (Figure~\ref{fig: allecc}). As noted earlier, upper mass limits for Styx and Kerberos are similar to the nominal masses in Table~\ref{tab: init}.  Estimates for Nix (Hydra) are 50\%--60\% (67\%) of the nominal masses.

To provide a preliminary analysis of the orbital stability of the two new state vectors, we began new sets of \nbody\ calculations. These calculations follow the same procedures as those described above. Here, we consider orbital elements for calculations spanning 100~Myr using the quoted masses from each study. Over this period, the orbital elements were stable, with no apparent oscillations as usually observed in unstable systems. However, the 100~Myr time scale is too short to judge whether these systems will reach 4.5~Gyr without any ejections. Eventually, we plan to report results of calculations covering 3--4.5~Gyr with a range of satellite masses. 

For each set of calculations, the measured Keplerian (osculating) orbital elements agree with the elements listed in \citet{porter2023} and \citet{brozovic2024}. As in Fig.~\ref{fig: allecc}, the geometric, corrected geometric, and iterative most circular $e$ and $\imath$ are systematically smaller than the Keplerian elements. Differences are largest for Styx and smallest for Hydra. In this respect, the new state vectors behave nearly identically to the HST state vector. 

\begin{figure}[t]
\begin{center}
\includegraphics[width=0.55\linewidth]{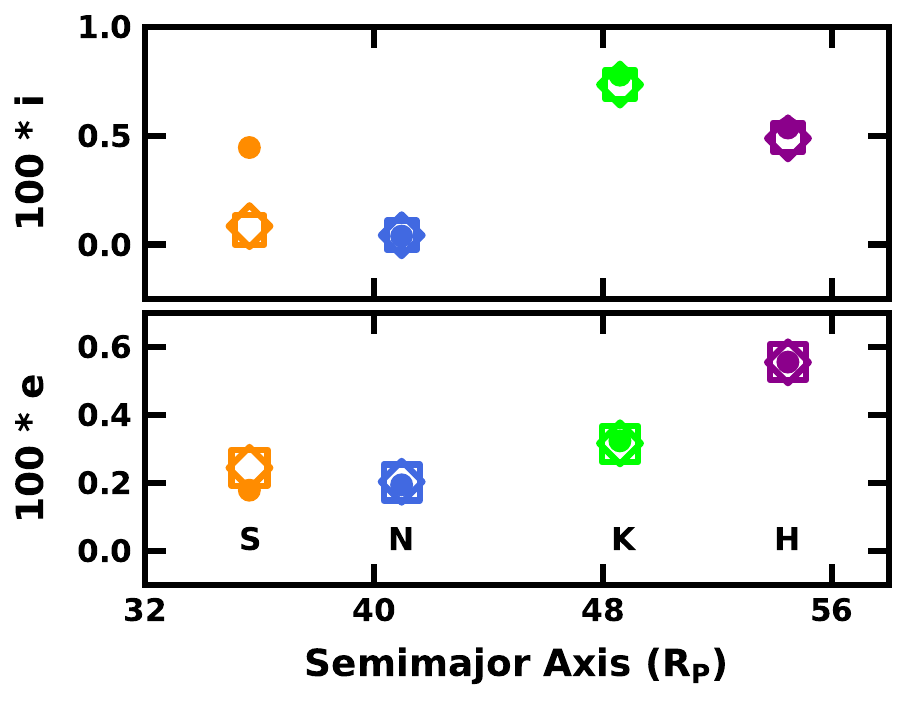}
\vskip -1ex
\caption{
\label{fig: aenew}
Comparison of the median orbital elements inferred from numerical calculations with the \citet{brozovic2015} state vector (filled circles) with similarly derived elements from the \citet[][open squares]{porter2023} and \citet[][open diamonds]{brozovic2024} state vectors. For Nix, Kerberos, and Hydra, $e$ and $\imath$ agree. For Styx, the numerical estimate for $e$ ($\imath$) from the new state vectors is 37\% larger (85\% smaller) than the estimate from the \citet{brozovic2015} state vector. The lower panel includes an initial for each satellite at $e$ = 0. 
}
\end{center}
\end{figure}

Figure~\ref{fig: aenew} compares estimates derived from the iterative most circular formalism for all three state vectors. Small filled circles are results for the HST state vector with $f$ = 0.5 listed in Table~\ref{tab: elements}. Larger open symbols are elements for the masses quoted in \citet{porter2023} and \citet{brozovic2024}. For Nix, Kerberos, and Hydra, calculations with the three state vectors yield nearly identical results for \efree\ and $\imath$. With the new state vectors, Kerberos and Hydra have somewhat smaller $\imath$; Nix has marginally larger \efree. 

Although the two new state vectors yield identical results for orbital elements of Styx, they are much different from the results for the HST state vector. Relative to the HST state vector, Styx has a 37\% larger \efree\ and a 85\% smaller $\imath$. With a larger \efree, Styx orbits closer to Nix at each apocenter with the new state vectors than with the HST state vector. The smaller $\imath$ in the new state vectors also places Styx closer to Nix at apocenter: gravitational pulls from Nix are then more in the orbital plane and generate larger radial motion than with the HST state vector.

Based on small-angle scattering theory, we hypothesize that the orbit of Styx in the new state vectors is somewhat more unstable than in the HST vector. In low mass \pc\ satellite systems, Styx is more sensitive to Nix than Kerberos is to Hydra (Figure~\ref{fig:evf}). In the small-angle scattering formalism we applied to understand this behavior with the HST state vector, the gravitational perturbations Styx receives from Nix scale as $\delta a_{SN}^{-5}$, where $\delta a_{SN}$ is the Styx--Nix separation at closest approach \citep[Styx apocenter;][]{bk2011b}. Compared to the situation with the HST state vector, Styx receives 3\% larger perturbations from Nix every synodic period of 107 days with the new state vector. Thus, we anticipate that Styx will be more prone to ejections. A complete suite of 3--4.5~Gyr \nbody\ calculations will test this preliminary conclusion about the stability of the two new state vectors. We anticipate completion in $\sim$ 2 yr.

\section{Discussion} \label{sec: disc}

Over the past 14 years, ground-based and satellite observations have revealed more than a dozen circumbinary planetary systems \citep[e.g.,][and references therein]{doyle2011,orosz2019,kostov2020,kostov2021,goldberg2023,standing2023}. Despite myriad challenges in deriving the properties of several systems, these studies identify systems with one or more terrestrial or gas giant planets orbiting a central binary. Aside from general fascination, these planets also provide a new laboratory to investigate dynamical stability and formation in systems where the gravitational potential varies with time \citep[e.g.,][]{lines2014,bk2015a,chavez2015,quarles2018b,trifonov2018,chen2019,childs2021,childs2023,kostov2023,langford2023,lubow2024}.  

Of all the known circumbinary planetary and satellite systems, \pc\ is by far the most accessible. With Styx, Nix, Kerberos, and Hydra, accurate positions yield high quality astrometric orbits. In principle, radial velocity data for the satellites would provide additional constraints on orbits and masses. Once these measurements become possible, comparisons between the \pc\ system and known circumbinary planetary systems could yield additional insights into the formation and dynamics of circumbinary systems.  

Here, we connect results from the \nbody\ calculations with other analyses of the \pc\ system and more distant circumbinary planetary systems. We focus on system masses and long-term stability, bulk densities, estimation of orbital parameters, and a discussion of work needed to make additional progress.
 
\subsection{System Masses and Long-term Stability}

The \nbody\ calculations discussed here and in \citet{kb2019b,kb2022} place strong upper limits on the masses of the small \pc\ satellites. For the nominal masses listed in Table~\ref{tab: init}, $\sim$ 53\% of the calculations reach 4.5~Gyr without ejecting a satellite. In contrast, satellite systems with $f \le$ 0.875 ($f \ge$ 1.25) are uniformly stable (unstable) over 4.5~Gyr. Based on these results, the \pc\ system is unlikely to have a mass larger than the sum of the nominal masses ($f \approx$ 1). Thus, we infer a robust upper limit $ m_{SNKH,max} \lesssim 9.5 \times 10^{19}$~g. Nearly all of this mass lies in Nix or Hydra; however it is possible to add or subtract mass from each of the satellites as long as the sum remains below $m_{SNKH,max}$.

These results are consistent with masses derived from analyses of HST and \nh\ astrometric data. \citet{porter2023} quote masses of $2.6 \pm 0.6 \times 10^{19}$~g (Nix), $3.0 \pm 0.3 \times 10^{19}$~g (Hydra), $ \le 8 \times 10^{17}$~g (Kerberos), and $\le 5 \times 10^{17}$~g (Styx). These masses roughly correspond to the masses considered here when $f$ = 0.625 (1.0) for Nix and Hydra (Styx and Kerberos). Considering the uncertainties in the \citet{porter2023} mass estimates, the 3-$\sigma$ upper limit for Nix (Hydra) is 98\% (80\%) of the nominal mass adopted in Table~\ref{tab: init}.  

\citet{brozovic2024} analyze the same data set and derive somewhat different orbital elements and masses than \citet{porter2023}. Aside from concluding that there are no constraints on the masses of Styx and Kerberos, they derive masses of $2.2 \pm 0.7 \times 10^{19}$~g (Nix) and $2.9 \pm 0.4 \times 10^{19}$~g (Hydra). These masses are 48\% (Nix) and 60\% (Hydra) of the nominal masses. The somewhat larger uncertainties in the masses result in 3-$\sigma$ upper limits that are 96\% (Nix) and 85\% (Hydra) of the nominal masses used here. 

In \S\ref{sec: newstate}, we use the first 100~Myr of numerical \nbody\ calculations with these new state vectors to estimate $a$, $e$, and $\imath$ for the small satellites. Although Nix, Kerberos, and Hydra have the same values for all three state vectors, Styx has larger $e$ and smaller $\imath$. With these orbital elements, Styx makes closer approaches to Nix every synodic period of 107 days and is thus more prone to ejection. If the \pc\ system has masses close to either of the 3-$\sigma$ upper limits for Nix and Hydra, it is marginally stable with the HST state vector. For either of the state vectors in \citet{porter2023} and \citet{brozovic2024}, the \nbody\ calculations with the HST state vector discussed here imply that extra gravitational forcing from Nix would result in many fewer intact systems with the nominal masses after 4.5~Gyr of dynamical evolution. While the impact on less massive systems is unclear, we anticipate that these systems would also be less stable. New sets of \nbody\ calculations are needed to verify these conjectures.

The results of the \nbody\ calculations are also consistent with calculations of closely packed planetary systems \citep[e.g.,][]{wisdom1980,petit1986,gladman1993,chambers1996,deck2013,fang2013,fabrycky2014,kratter2014,mahajan2014,pu2015,morrison2016,obertas2017,weiss2018,sutherland2019}. In these calculations, ensembles of a few planets are unstable when their separations in Hill units are $K \le$ 8--10. For the  \pc\ satellite system, $K_{SN} = 12/f^{1/3}$, $K_{NK} = 16/f^{1/3}$, and $K_{KH} = 10/f^{1/3}$. The nearly 100\% failure rate for systems with $f$ = 1.25 and the 40\% to 50\% failure rate for systems with $f$ = 1 agrees with the larger set of \nbody\ calculations for exoplanetary systems. The negligible failure rate for \pc\ satellite systems with $f$ = 0.875 similarly follows.


\subsection{Bulk Density}

Together with imaging data from \nh, the masses derived in \citet{porter2023}, \citet{brozovic2024}, and here constrain the bulk densities of the small \pc\ satellites. \citet{kb2019b} adopt a triaxial model to estimate volumes from the size measurements of \citet{weaver2016}. Converting to equivalent spherical radii yields 5.2~km (Styx), 19.3~km (Nix), 6~km (Kerberos), and 20.9~km (Hydra). Analysis of the full set of \nh\ imaging data yields radii of 5.25~km (Styx), 18.25~km (Nix), 6~km (Kerberos), and 18.1~km \citep[Hydra;][]{porter2021}. The two approaches agree for Styx and Kerberos; the more detailed study in \citet{porter2021} results in a 16\% (35\%) smaller volume for Nix (Hydra) and a larger bulk density.

For the masses adopted above ($f \approx$ 1) and volumes from \citet{porter2021}, we infer bulk densities $\rho_S \lesssim 1.01$~\gcmc\ (Styx), $\rho_N \approx 1.71$~\gcmc\ (Nix), $\rho_K \lesssim 1.11$~\gcmc\ (Kerberos), and $\rho_H \approx$ 1.70~\gcmc\ (Hydra). Adopting the 3-$\sigma$ upper limits on the masses in \citet{porter2023} and \citet{brozovic2024} implies similar results. Due to the smaller volumes, the bulk densities for Nix and Hydra are larger than those estimated in \citet{kb2019b}.

Adopting smaller masses for Nix and Hydra imply correspondingly lower bulk densities. A robustly stable satellite system with $f$ = 0.5 has $\rho_S \lesssim 0.5$~\gcmc, $\rho_N \approx 0.85$~\gcmc, $\rho_K \lesssim 0.55$~\gcmc, and $\rho_H \approx 0.85$~\gcmc. The mass estimates in \citet{porter2023} result in $\sim$ 20\% (44\%) larger bulk densities for Nix (Hydra). \citet{brozovic2024} infer $\rho_N \approx 0.88$~\gcmc\ and $\rho_H \approx 1.21$~\gcmc. Within the errors, all of these bulk density estimates are consistent with an upper limit of $\sim$ 1.2~\gcmc\ for all of the small satellites. 

To estimate the most likely bulk densities for Nix and Hydra, \citet{kb2019b} performed a Monte-Carlo calculation that accounted for uncertainties in the masses and sizes. From this analysis, the probability that the satellites have smaller bulk densities than Charon (Pluto) is 65\% (80\%) for Nix and 75\% (90\%) for Hydra. Hydra's larger volume in \citet{kb2019b} is responsible for the higher probabilities. With the smaller volumes for Nix and Hydra from \citet{porter2023}, the nominal bulk densities of Nix and Hydra for $f$ = 1 are then indistinguishable from the bulk density of Charon 
($\rho_C = 1.705$~\gcmc)
and only slightly smaller than the bulk density of Pluto \citep[$\rho_P$ = 1.853~\gcmc;][]{brozovic2024}. Adopting the 3-$\sigma$ upper limits on satellite masses from the HST + \nh\ astrometric analyses yields similar results. 

If the \pc\ satellites have masses similar to the nominal masses with $f$ = 0.5 or to those derived in \citet{porter2023} and \citet{brozovic2024}, then their densities are very likely smaller than those of Charon and Pluto. Repeating the same Monte Carlo analysis in \citet{kb2019b}, Nix has a better than 95\% (98\%) probability that its density is smaller than Charon (Pluto). The smaller volume of Hydra relative to that of Nix implies lower probabilities of 76\% (91\%) that the bulk density is smaller than Charon (Pluto). In addition to its smaller volume, the larger errors in the volume contribute to the lower probability that the bulk density of Hydra is smaller than either Pluto or Charon.

These bulk density estimates place some constraints on the porosity of Nix and Hydra. For the nominal masses ($f$ = 1) and the measured volumes, Nix and Hydra have bulk densities similar to (slightly smaller than) the bulk density of Charon (Pluto). Unless the small satellites are composed of material denser than either Pluto or Charon, Nix and Hydra must have a small porosity, $\lesssim$ 8\%. A satellite system with $f \approx$ 0.5 provides an upper limit on porosity, $\sim$ 50\%, for a composition similar to Charon or Pluto. The masses derived in \citet{porter2023} and \citet{brozovic2024} imply a porosity between these two extremes: $\sim$ 40\% for Nix and $\sim$ 30\% for Hydra. 

Small satellites of Uranus and Neptune have bulk densities similar to that of Nix and Hydra \citep[e.g.,][]{chancia2017,brozovic2020,french2024}. The bulk densities of Cressida, Ophelia, and Cordelia, $\rho \approx$ 0.7--1.4~\gcmc, closely follow the variation of the Roche limit with distance from Uranus. The two innermost moons of Neptune, Naiad and Thalassa, have $\rho \approx$ 0.8--1.2~\gcmc. Unlike Nix and Hydra, the masses of the Uranian moons are more accurately established by their impact on the ring system. However, their volumes are somewhat uncertain. The two Neptunian moons have comparable uncertainties in mass and volume relative to Nix and Hydra.

Saturn's ring moons also have high porosity. Atlas, Daphnis, Epimetheus, Janus, Pan, Pandora, and Prometheus have bulk densities in the range 0.27--0.64~\gcmc\ derived from measured radii and dynamical mass estimates \citep{jacobson2004,renner2005,cooper2015,ciarniello2024}. If composed of water ice with a typical density $\sim$ 1~\gcmc, these moons have high porosity, $\sim$ 50\%, similar to Nix and Hydra when $f \approx$ 0.5--0.6. Small Saturnian moons also have high albedo, $\sim$ 0.5 \citep{ciarniello2024}, comparable to the albedo measured for all four of the small \pc\ satellites \citep{weaver2016,cook2018}.

\citet{ciarniello2024} review two plausible formation mechanisms for small Saturnian moons \citep{canup1995,charnoz2007,porco2007}: (i) accretion of ices by a denser seed within the rings and (ii) spontaneous accretion of icy material at or near the Roche limit. Although the small \pc\ satellites are well outside the Roche limit, agglomeration of ices around larger icy or rocky particles is possible within material (i) left over from the giant impact that produced the central binary \citep[e.g.,][]{walsh2015,kb2021a} or (ii) ejected during a collision between Charon and a wayward trans-Neptunian object \citep{bk2020}. Either approach would generate low density, high porosity, and high albedo objects in a manner similar to processes within Saturn's rings. 

\subsection{Orbital Elements}

Circumbinary orbits are not simple ellipses \citep{lee2006,leung2013,bk2015a,woo2020,bk2021,gakis2023,langford2023}. On `circular' orbits, objects feel a time-variable gravitational potential that evolves with the orbital period, $P_b$, of the central binary. For circular binaries like \pc, analytical and numerical methods characterize satellite orbits rather well. Far from the system barycenter, ellipses provide reasonable matches to orbits and allow straightforward estimation of orbital elements and their errors. For orbital periods $\lesssim 10 ~ P_b$, however, the analysis becomes more complicated. Several approaches lead to acceptable results. 

When circumbinary orbits are well-sampled on time scales comparable to the orbital period, $P$, we recommend the \citet{woo2020} FFT method. Coupled with the analytical theories of \citet{lee2006} and \citet{leung2013}, this approach yields accurate circumbinary orbital elements and enables robust error estimation. Although the technique requires some extra computation, the improved accuracy for Styx and Nix is worth the effort.

When orbits are well-sampled on much longer time scales, orbital precession limits the use of the FFT and similar solutions \citep[e.g.,][]{woo2020,langford2023}. We then recommend one of the geometric approaches outlined in \S\ref{sec: analysis} above. As summarized in Figure~\ref{fig: allecc}, elements inferred from instantaneous measurements of position and velocity yield more accurate results than Keplerian estimates derived from the instantaneous velocity and angular momentum. 

\subsection{Future Prospects}

With the stability of the HST state vector from \citet{brozovic2015} now ensured, additional \nbody\ calculations of this system of four satellites are not useful. With new state vectors from the combined HST + \nh\ astrometry, stability calculations with an \nbody\ code would enable constraints on the upper mass limit on a stable satellite system and the ability of these systems to maintain stability for $\gtrsim$ 4.5~Gyr. We have started such calculations and illustrate preliminary results in Figure~\ref{fig: allecc}. We plan to report complete results in a future publication.

We envision two lines of research to improve the mass estimates of Styx, Nix, Kerberos, and Hydra. Currently, JWST images can easily detect these four satellites with sufficient signal-to-noise to pinpoint their positions relative to the orbits derived from HST + \nh\ images \citep[e.g.,][]{kb2019a}. Once the satellites are tracked for many orbits, new orbital solutions should provide better constraints on the orbital parameters and satellite masses. Given the improvements made possible with \nh\ data, additional JWST data might eliminate the possibility that the satellites could have masses close to the nominal masses in Table~\ref{tab: init}. This outcome could ensure that the satellite system is robustly stable for main sequence lifetime of the Sun. 

Independently of orbital fits, JWST imaging data can constrain the mass of Hydra in a fashion similar to the methods used to estimate masses for moons close to the ring system of Uranus \citep{french2024}. \citet{kb2019a} demonstrated that small satellites outside the orbit of Hydra are stable at orbital distances $\gtrsim$ 1.15~$a_H$, which is $\sim$ 9700~km beyond Hydra's semimajor axis $a_H \approx$ 65000~km \citep{brozovic2015,porter2023,brozovic2024}. For the nominal masses in Table~\ref{tab: init}, 9700~km is $\sim$ 14.5~$R_{H,H}$ where $R_{H,H}$ is the Hill radius of Hydra. If the mass of Hydra is smaller, $\sim$ half of the nominal mass, the 25\% smaller $R_{H,H}$ implies that the orbits of small satellites are stable at $\gtrsim$ 7750~km beyond Hydra's orbit instead of 9700~km. If another satellite or dusty material lies within 7,000--10,000~km of Hydra's orbit, this discovery would enable a new upper limit on the mass for Hydra and provide tests of masses derived from orbital fits to HST + \nh\ astrometric data.

\section{Summary} \label{sec: summary}

We analyze a new suite of $\sim$ 275 numerical \nbody\ calculations with initial conditions based on the HST state vector derived in \citet{brozovic2015}. These results extend previous studies on orbital architecture \citep{kb2019a}, masses \citep{kb2019b}, and the bulk density and evolutionary behavior of the small circumbinary satellites \citep{kb2022}. Here, we focus on final estimates for satellite masses and orbital elements and on the long-term stability of the \citet{brozovic2015} state vector. Together, the full suite of \nbody\ calculations allow us to make the following conclusions; we refer readers to the published papers for more details.

\begin{itemize}

\item 
To make progress, we adopt a set of nominal masses (Table~\ref{tab: init}) and perform \nbody\ calculations with masses that are a factor $f_i$ times the nominal mass, where $i$ refers to Styx, Nix, Kerberos, or Hydra. When all satellites have the same $f_i$, we use $f$ to denote the mass factor. In \citet{kb2019b}, the calculations show that a `heavy' system with masses close to those adopted in \citet[][$f_S$ = 7.5, $f_H$ = 1, $f_K$ = 16.5, and $f_H$ = 1]{brozovic2015} is unstable on time scales $\lesssim$ 1~Gyr. In contrast, the `light' system with the masses listed in Table~\ref{tab: init} ($f_i$ = 1 for all satellites) is stable on much longer time scales.

\item
The full suite of \nbody\ calculations demonstrates that all satellite systems with $f \ge$ 1.5 are unstable on time scales inversely proportional to $f$ \citep[here; see also][]{kb2019b,kb2022}. These calculations always eject at least one satellite, either Styx or Kerberos, and sometimes Hydra. Nearly all calculations with $f$ = 1.25 are unstable; several maintain stability for the 4.5~Gyr duration of an \nbody\ calculation. In contrast, roughly half of the calculations with $f$ = 1.0 and all calculations with $f \le$ 0.875 are stable on 4.5~Gyr time scales (see Table~\ref{tab: lifetimes}). 

\item
Based on the stability of simulated satellite systems, we establish an upper limit on the total mass in the four small satellites, $m_{SNKH,max} \lesssim 9.5 \times 10^{19}$~g \citep[here; see also][]{kb2019b,kb2022}. For this total mass, the masses of Styx and Kerberos are very unlikely to be more than 2--3 times their nominal masses. Although we prefer the nominal masses for Nix and Hydra, somewhat larger masses for either satellite are allowed if the total mass of the four satellites remains less than or comparable to $m_{SNKH,max}$. Satellite sizes deduced from \nh\ data yield an estimate for the bulk densities: $\rho_S \lesssim 1.01$~\gcmc, $\rho_N \approx 1.71$~\gcmc, $\rho_K \lesssim 1.11$~\gcmc, and $\rho_H \approx$ 1.70~\gcmc.

\item
Satellite systems with $f \lesssim$ 0.875 ($m_{SNKH} \lesssim 8 \times 10^{19}$~g) maintain nearly constant $e$ and $\imath$ over the current age of the Sun, $\sim$ 4.5~Gyr. The two mass estimates for the satellite system are consistent with the 3-$\sigma$ upper limits on system masses derived from astrometric analyses of HST and \nh\ imaging data \citep{porter2023,brozovic2024}. Smaller masses for Nix and Hydra imply lower bulk densities, $\rho_{NH} \sim$ 0.8--1.2~\gcmc, that are similar to the bulk densities of the Uranian moons Cressida, Ophelia, and Cordelia \citep{french2024}, the Neptunian moons Naiad and Thalassa \citep{brozovic2020}, and the Saturnian moons Atlas, Daphnis, Epimetheus, Janus, Pan, Pandora, and Prometheus \citep[][and references therein]{ciarniello2024}. The good agreement between the bulk densities of these satellites points to lower masses (e.g., $f$ = 0.50--0.625) for Nix and Hydra.

\item
The 1000~yr \nbody\ calculations discussed above illustrate that Kerberos is more sensitive than Styx to increases in the total mass of the satellite system. When $f$ = 3, the eccentricity of Kerberos after 1000~yr of evolution is roughly twice the eccentricity for calculations with $f \approx$ 0.0--0.5. In contrast, Styx' eccentricity hardly changes as $f$ grows from 0.0--0.5 to 3.0. This difference in the initial response of Styx and Kerberos to changes in satellite masses explains the ejection statistics outlined here and in \citet{kb2022}, where Kerberos (Styx) ejections are much more likely in satellite systems with $f \gg 1$ ($f \approx$ 1). In systems with large $f$, Kerberos is more easily excited to large $e$, which facilitates resonance crossing and subsequent ejection. In systems with $f \approx$ 1, Styx's orbit is more easily excited on time scales longer than 1--10~Myr and is therefore ejected more often than Kerberos. In these situations, Styx is more likely to `signal' an impending ejection; after Styx's $\imath$ grows beyond 0.01, the inclination and eccentricity may remain quasi-stable for several tens or hundreds of Myr prior to ejection.

\item 
When the satellites have their nominal masses, massless particles with nearly circular orbits at distances ranging from just inside the orbit of Styx to just outside the orbit of Hydra have short lifetimes, $\sim$ 100--1000~yr up to 10--100~Myr \citep{kb2019a}. The four small satellites drive  nearby massless particles onto orbits that cross resonances with the central binary. After resonance crossing, massless particles find their way inside the innermost stable orbit of the binary and then are ejected. The time scale to clear material from the vicinity of the small satellites scales with their mutual Hill radii. Although a lower mass satellite system with $f \approx$ 0.50--0.625 would then take somewhat longer to clear material from the orbits of the four small satellites, the time scale is still shorter than the age of the Sun. Massless particles orbiting at distances $\gtrsim$ 1.1~$a_H$, where $a_H$ is the semimajor axis of Hydra's orbit, are stable on 10--100~Myr time scales. These particles are also safe from the gravities of the Sun and major planets \citep{michaely2017}; thus, their orbits are likely to remain stable for the main sequence lifetime of the Sun. The short lifetimes of particles orbiting between Styx and Hydra explains the lack of dust detected during the \nh\ mission \citep{lauer2018}.

\item
For any circumbinary orbit close to the central binary, we recommend using the methods outlined in \S\ref{sec: analysis} to derive the eccentricity \citep[see also][]{bk2021,kb2022}. While elements derived from fitting ellipses to orbits \citep[e.g.,][]{showalter2015} agree with these dynamical estimates for Kerberos and Hydra, they fail for Nix and Styx (Figure~\ref{fig: ae}). During long \nbody\ integrations, the most-circular parameters, \rgc\ and \efree, yield more accurate results for the semimajor axis and eccentricity than Keplerian (osculating) estimates derived from the instantaneous energy and angular momentum. Geometric elements (eqs.~\ref{eq: ag} and \ref{eq: eg}) provide good first estimates of the semimajor axis and eccentricity that can be improved by iteration. When many measurements are available throughout 1--2 circumbinary orbits, the FFT method of \citet{woo2020} is superior to all these methods. 

\end{itemize}

Improving the available mass estimates is challenging. If small satellites with radii $\sim$ 1--2~km lie outside the orbit of Hydra, JWST imaging data can detect them \citep{kb2019a}. These satellites can be no closer to the system barycenter than a few mutual Hill radii outside Hydra's orbit. Precise measurements of the innermost orbital distance of this material would constrain the Hill radius and thus the mass of Hydra. If Nix and Hydra have similar compositions, a better estimate for Nix' mass follows from the ratio of volumes. Larger uncertainties in the volumes of Styx and Kerberos probably preclude improving their masses in this way.

JWST astrometric data can also improve the orbits of the four known satellites \citep{kb2019a}. Analyses of the HST + \nh\ data reduced uncertainties in the masses of Nix and Hydra by factors of 2--3. Similar reductions seem possible with JWST. 
If these are achieved, comparisons between the results of the \nbody\ calculations discussed here and a real system with more accurate system masses would allow a better understanding of the evolution of the \pc\ satellite system. 

\vskip 6ex

We thank two anonymous referees whose cogent comments helped us hone our presentation.  We acknowledge generous allotments of computer time on the NASA `discover' cluster, provided by the NASA High-End Computing (HEC) Program through the NASA Center for Climate Simulation (NCCS). Advice and comments from M. Geller improved our discussion. Portions of this project were supported by the {\it NASA } {\it Outer Planets} and {\it Emerging Worlds} programs through grants NNX11AM37G, NNX17AE24G, and 80NSSC23K0252. Some of the data (binary output files and C programs capable of reading them) generated from this numerical study of the \pc\ system are available at a publicly-accessible repository (https://hive.utah.edu/) with url https://doi.org/10.7278/S5d-0fwz-1gny.

In addition to the analyses described here and in \citet{kb2019b,kb2022}, we deposit binary files from all completed \nbody\ calculations and some of the programs used to extract and analyze the phase space coordinates at a publicly accessible repository (https://hive.utah.edu/). The combined set of 700 files from \citet{kb2019b}, the 500 files from \citet{kb2022}, and 275 files from this study provides interested researchers a large data set for other analyses.

\bibliography{ms.bbl}{}
\bibliographystyle{aasjournal}

\end{document}